\documentclass[a4paper,10pt]{article}
\usepackage[utf8x]{inputenc}
\usepackage[]{graphicx}

\usepackage{fancyhdr}
\title{\bf ASTROCLADISTICS: 
MULTIVARIATE EVOLUTIONARY ANALYSIS IN ASTROPHYSICS}
\author{\bf Didier Fraix-Burnet \\
\small Université Joseph Fourier - Grenoble 1 / CNRS \\
\small Laboratoire d'Astrophysique de Grenoble (LAOG) UMR 5571 \\
\small BP 53, F-38041 GRENOBLE Cedex 09, France \\
\small Email: fraix@obs.ujf-grenoble.fr}
\date{}

\begin{document}

\maketitle
\thispagestyle{fancy}
\fancyhead[L]{Proc. ICCS-X, Cairo, Egypt December 20-23, 2009, Vol. 18, pp. xx-xx}
\fancyhead[R]{}

\begin{abstract}
        The Hubble tuning fork diagram, based on morphology and established in the 1930s, has always been the preferred scheme for classification of galaxies. However, the current 
large amount of data up to higher and higher redshifts asks for more sophisticated statistical approaches like multivariate analyses. Clustering analyses are still very confidential, and do not take into account the unavoidable characteristics in our Universe: evolution. Assuming branching evolution of galaxies as a 'transmission with modification', we have shown that the concepts and tools of phylogenetic systematics (cladistics) can be heuristically transposed to the case of galaxies. This approach that we call ``astrocladistics'', has now successfully been applied on several samples of galaxies and globular clusters. Maximum parsimony and distance-based approaches are the most popular methods to produce phylogenetic trees and, like most other studies, we had to discretize our variables. However, since astrophysical data are intrinsically continuous, we are contributing to the growing need for applying phylogenetic methods to continuous characters.

        classification; astrophysics; cladistics; multivariate; evolution; galaxies; continuous characters

\end{abstract}

\section{INTRODUCTION}

        The extragalactic nature of galaxies was discovered by Edwin Hubble less than 90 years ago. He also discovered the expansion of the Universe and established in 1936 the famous tuning fork diagram or Hubble diagram based on the idea that elliptical galaxies should evolve into flattened systems such as ordinary or barred spiral galaxies. Since then, telescopes and the associated detectors have made remarkable progress. We are now able to study galaxies in great detail, identifying individual stars, gas and dust clouds, as well as different stellar populations. Imagery brings very fine structural details, and spectroscopy provides the kinematical, physical and chemical conditions of the observed entities at different locations within the galaxy. For more distant objects, information is scarcer, but deep systematic sky surveys gather millions of spectra for millions of galaxies at various redshifts.

        Like paleontologists, we observe objects from the distant past (galaxies at high redshift), and like evolutionary biologists, we want to understand their relationships with nearby galaxies, like our own Milky Way. Strangely enough, the Hubble classification, depicted in the Hubble diagram, is still frequently used as a support to describe galaxy evolution, even though it ignores all observables except morphology (Fraix-Burnet et al 2006a, Hernandez and Cervantes-Sodi 2006). Can the enormous amount of very detailed observations and galaxy diversity be solely depicted by few large families characterized only by their global shape? There obviously must be a better way to exploit the data gathered by very large telescopes and their sophisticated detectors, and to account for the complicated physical processes that lead to the diversification of galaxy.

        Abandoning the one-parameter classification approach and using all available descriptors means taking a methodological step equivalent to the one biologists took after Adanson and Jussieu in the 18th century. One basic tool, the Principal Component Analysis, is relatively well-known in astrophysics (e.g. Cabanac et al 2002, Recio-Blanco 2006). However, only a very few attempts to apply multivariate clustering methods have been made very recently (Chattopadhyay and Chattopadhyay 2006, 2007; Chattopadhyay et al 2007, 2008, 2009). Sophisticated statistical tools are used in some areas of astrophysics and are developing steadily, but multivariate analysis and clustering techniques have not much penetrated the community.

        A supplementary difficulty is that evolution, an unavoidable fact, is not correctly taken into account in most classification methods. By mixing together objects at different stages of evolution, most of the physical significance and usefulness of the classification is lost. In practice, the evolution of galaxies is often limited to the evolution of the properties of the entire population as a function of redshift (Bell 2005). Since environment (the expanding Universe) and galaxy properties are so much intricate, this kind of study is relevant to a first approximation. However, recent observations have revealed that galaxies of all kinds do not evolve perfectly in parallel, as illustrated for instance by the so-called downsizing effect which shows that large galaxies formed their stars earlier than small ones (e.g. Neistein 2006). New observational instruments now brings multivariate information at different stages of evolution, and in various evolutive environments. In this multivariate context, we believe that the notion of "evolution", easy to understand for a single parameter, is advantageously replaced by "diversification".

        In this conference, I present our efforts to implement a method to reconstruct the "galactogenesis", based on tools and concepts largely developed in evolutionary biology and bioinformatics. I first present cladistics and some of our results. Then I replace this approach among multivariate statistical methods and explain why we also need these other tools to explore the parameter space. Finally, I present one of our developments that makes the astrophysical problematics to join current bioinformatics and mathematics studies about the use of continuous characters in reconstructing phylogenies.

\section{ASTROCLADISTICS}

Multivariate clustering methods compare objects with a given measure and then gather them according to a proximity criterion. Distance analyses are based on the overall similarity derived from the values of the parameters describing the objects. The choice of the most adequate distance measure for the data under study is not unique and remains difficult to justify a priori. The way objects are subsequently grouped together is also not uniquely defined. Cladistics uses a specific measure that is based on characters (a trait, a descriptor, an observable, or a property, that can be given at least two states characterizing the evolutionary stages of the object for that character) and compares objects in their evolutionary relationships (Wiley 1991). Here, the “distance” is an evolutionary cost. Groupings are then made on the basis of shared or inherited characteristics, and are most conveniently represented on an evolutionary tree.

        Character-based methods like cladistics are better suited to the study of complex objects in evolution, even though the relative evolutionary costs of the different characters is not easy to assess. Distance-based methods are generally faster and often produce comparable results, but the overall similarity is not always adequate to compare evolving objects. In any case, one has to choose a multivariate method, and the results are generally somewhat different depending on this choice (e.g. Buchanan 2008). However, the main goal is to reveal a hidden structure in the data sample, and the relevance of the method is mainly provided by the interpretation and usefulness of the result. 

                Multivariate evolutionary classification in astrophysics has been pioneered by the author (Fraix-Burnet et al 2006a, 2006c, Fraix-Burnet 2009). Called astrocladistics, it is based on cladistics that is heavily developed in evolutionary biology. Astrocladistics has been first applied to galaxies (Fraix-Burnet et al 2006b) because they can be shown to follow a transmission with modification process when they are transformed through assembling, internal evolution, interaction, merger or stripping. For each transformation event, stars, gas and dust are transmitted to the new object with some modification of their properties. Cladistics has also been applied to globular clusters (Fraix-Burnet et al 2009), where interactions and mergers are probably rare. These are thus simpler stellar systems, even though we have firm evidence that internal evolution can create another generation of stars and that globular clusters can lose mass. Basically, the properties of a globular cluster strongly depend on the environment in which it formed (chemical composition and dynamics), and also on the internal evolution which includes at least the aging of its stellar populations. Since galaxies and globular clusters form in a very evolving environment (Universe, dark matter haloes, galaxy clusters, chemical and dynamical environment), the basic properties of different objects are related to each other by some evolutionary pattern. 

 \begin{figure}[t]
         \begin{center}
                 \includegraphics[width=10 true cm,angle=0]{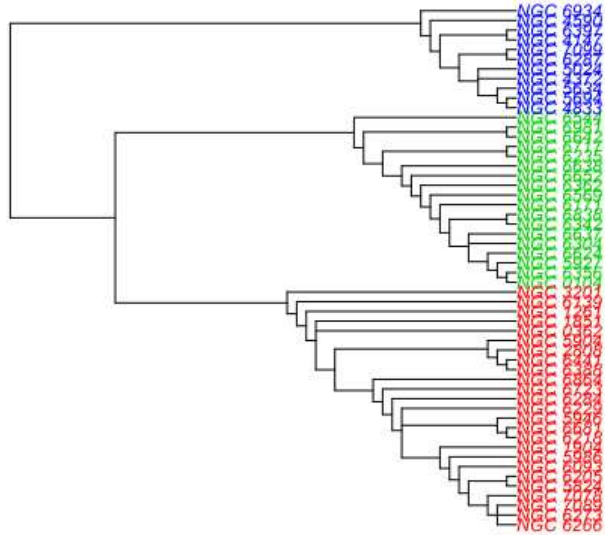}
 \caption{Cladogram of Galactic Globular Clusters.}
         \label{fig1}
         \end{center}
 \end{figure}

 \begin{figure}[t]
         \begin{center}
                 \includegraphics[width=10 true cm,angle=0]{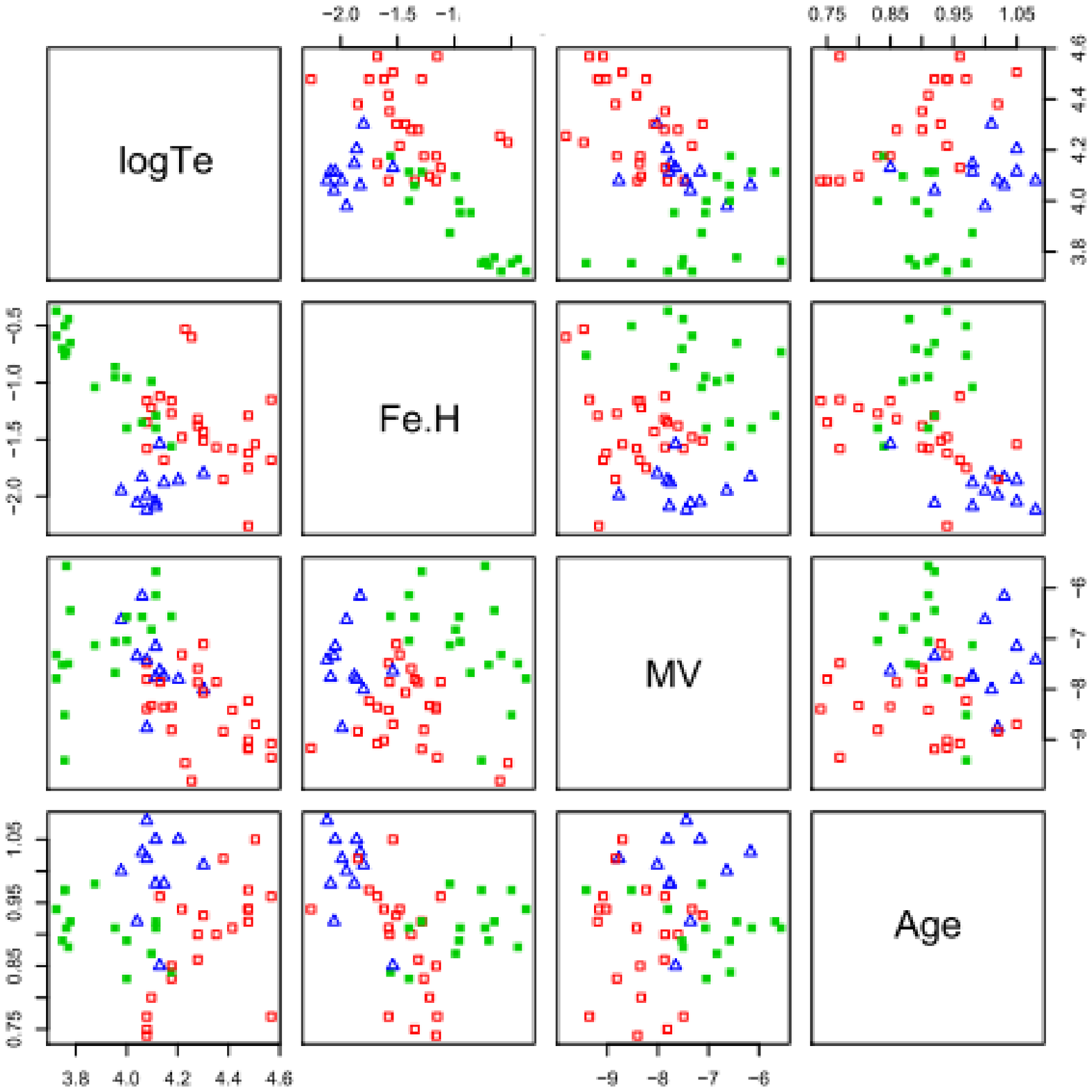}
 \caption{Scatterplots  for the 4 parameters used in the analysis, colors corresponding to groups defined on the cladogram of Figure 1.}
         \label{fig2}
         \end{center}
 \end{figure}

        In both cases, there is a kind of transmission with modification process, which justifies a priori the use of cladistics. It must be clear that this is not a "descent with modification" in the sense that there is no replication. But evolution does nevertheless create diversity. We are dealing with phylogeny (relationships between species), not with genealogy (relationships between individuals). Since a multivariate classification of galaxies is not yet available, we assume that each object represents a “species” that will have to be defined later on. The word ``progenitor'', often used for galaxy evolution, is to be understood in this way.

        A cladistic analysis works as follows. One first builds a matrix with values of the four parameters for all galaxies. In contrast to multivariate distance methods, undocumented values are not a problem in cladistic analyses. The values for each parameter are discretized into 30 bins representing supposedly evolutionary states. Discretization of continuous variables is quite a complex problem, especially in the evolutionary context (see Section 4). Here, we took equal-width bins. The choice of the number of bins cannot be made in a simple objective way. A priori, one could consider a compromise between an adequate sampling of continuous variables and the uncertainties on the measurements. The first constraint is given by the software (32 in this case). The second one would a priori give a lower limit of something like total range/uncertainty, but Shannon's theorem would multiply this by 2. Hence, 30 bins would account for about 7\% measurement errors. Even so, border effects always imply that some objects could belong to a bin or its neighbor, a process that add some more artificial noise. The best way to avoid this effect is to make several analyses with different number of bins and check that the result does not depend on this number. From our experience in astrocladistics, we know that the results are very generally identical between 20 and 30, and can differ with 10 bins. For lower number of bins, the results are very dependent on this number.

        Then, all possible arrangements of galaxies on a tree-like structure are constructed, and using the discretized matrix, the total number of state changes is computed for each tree. The most parsimonious tree is finally selected. If several such trees are found, then a consensus (strict or majority rule) tree is built. The whole procedure is computerized since the number of arrangements is very large. The result (depicted on a cladogram) is a diversification scenario that should be confronted to other knowledge and parameters. 

        Figure 1 shows the cladogram obtained for globular clusters of our Galaxy (Fraix-Burnet et al 2009). Three groups are identified. The first one (in blue) has on average the lower ratio Fe/H that measures the proportion of heavy atomic elements that are processed within stars. This group is consequently considered as more primitive.

        Figure 2 shows bivariate diagrams for the 4 parameters used in the analysis: logTe, that measures the temperature of stars that are at a specific point in their evolution,  Fe/H, MV that is the total visible intensity (magnitude) and roughly indicates the mass of the globular cluster, and Age that can be measured quite precisely because all stars of a given globular cluster are formed nearly at the same time. However Age is not an intrinsic property discriminating evolutionary groups since it evolves in the same way for all. But we gave it a half weight to arrange the objects within each group. Looking at plots like Figure 2 and using other parameters (orbital elements, kinematics, more refined chemical abundances...), we are able to infer that each group formed during a particular stage of the assembly history of our Galaxy. The blue group is the older one. It formed during the dissipationless collapse of the protogalaxy. They are located mainly in the outer halo. The red group belongs to the inner halo and the corresponding clusters formed at a later stage during the dissipational phase of Galactic collapse, which continued in the halo after the formation of the thick disc and its globular clusters. These clusters were very massive before "star evaporation" took place. The latter group (green) formed during an intermediate and relatively short period and comprises clusters of the disk of our Galaxy.

\section{STRUCTURE OF THE PARAMETER SPACE}

        The spectacular astrophysical results obtained so far were possible because the characters are pertinent from the evolutionary and physical points of view. However, they are too few to obtain a detailed classification of tens of thousands of galaxies. Since astrophysics is a science of observation, not of experimentation, it is difficult to obtain the intrinsic physical and chemical properties of galaxies or globular clusters especially if distant. The spectra carries all the information we gather, so that much data is available. But this information is generally not straightfully pertinent, either because it is not directly connected to physical parameters, or because there are redundancies or incompatibilities that are paticularly annoying in a cladistic analysis. Being as multivariate as possible ensures objectivity, but some care is required in selecting the parameters. Different multivariate tools are thus needed to explore the structure of the data.

        The first and obvious approach is to use Principal Component Analysis (PCA) to reduce the dimensionality of the parameter space and avoid redundancy. It is possible to perform a cladistic analysis directly with the main components. However, the correlations revealed by such analyses are not necessarily intrinsic nor due to scaling effects. They can be generated by evolution. For instance, in Figure 3, the parameter logs, that measures the central velocity dispersion, is not physically linked to mgbfe, that measures to the global metallic composition of the stars, but they appear "correlated" through their respective evolutions. They possibly characterize two independent facets of galaxy diversification. Such information would be lost with the use of the PCA components.

        Alternatively, it is possible to select the parameters that appear to have more weight in the PCA. In this approach, we both keep the evolutionary information and gain a precious knowledge on the parameters themselves, which are physical in contrast to the principal components.  Parameters for subsequent analyses are thus selected according to their loadings and their physical meaning. At the end, we eliminate true redundancies (parameters that measure the same quantity) and undiscriminating parameters.

 \begin{figure}[t]
         \begin{center}
                 \includegraphics[width=10 true cm,angle=0]{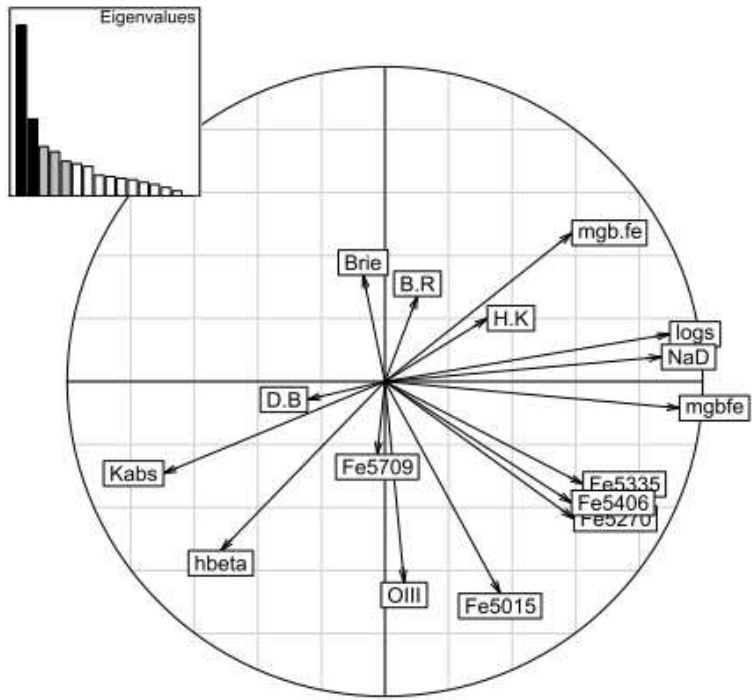}
 \caption{Projection on the two first Principal Components of parameters available for a typical sample of galaxies.}
         \label{fig3}
         \end{center}
 \end{figure}

        There are other techniques that surely could be interesting and that we have not yet explored. Rather, we are investigating approaches more specific to the problem of obtaining a phylogeny with continuous characters (see Section 4). It appears that the structure of the parameter space must be at least grossly understood before embarking on sophisticated clustering tools, especially if the evolutionary information is to be retrieved. One difficulty in multivariate clustering of continuous and evolutive parameters, certainly not specific to astrophysics, is the cosmic variance that makes the groups to largely overlap in the parameter space.

        Once the parameters are selected, it is instructive to compare results obtained with several multivariate clustering techniques in several different conditions (sub-samples, subsets of parameters). This provides additional insights on the parameters, and the different groupings can be confronted and analysed. Similarities reinforce the results, and mismatches can be interpreted in light of specificities of the techniques. In this way, we have obtained very convincing results on samples of hundreds of galaxies (Fraix-Burnet, Chattopadhyay, Chattopadhyay, Davoust and Thuillard, in prep). We demonstrate that multivariate clustering techniques and cladistics are the right tools to use. They yield results that are largely consistent, the cladogram providing the additional evolutionary relationshups between the groups. These groups cannot be guessed by any a priori splitting of the data, but we find that their properties happen to delineate different assembling histories, like for the globular clusters above.

\section{CONTINUOUS CHARACTERS AND SPLIT NETWORKS}

 \begin{figure}[t]
         \begin{center}
                 \includegraphics[width=6 true cm,angle=0]{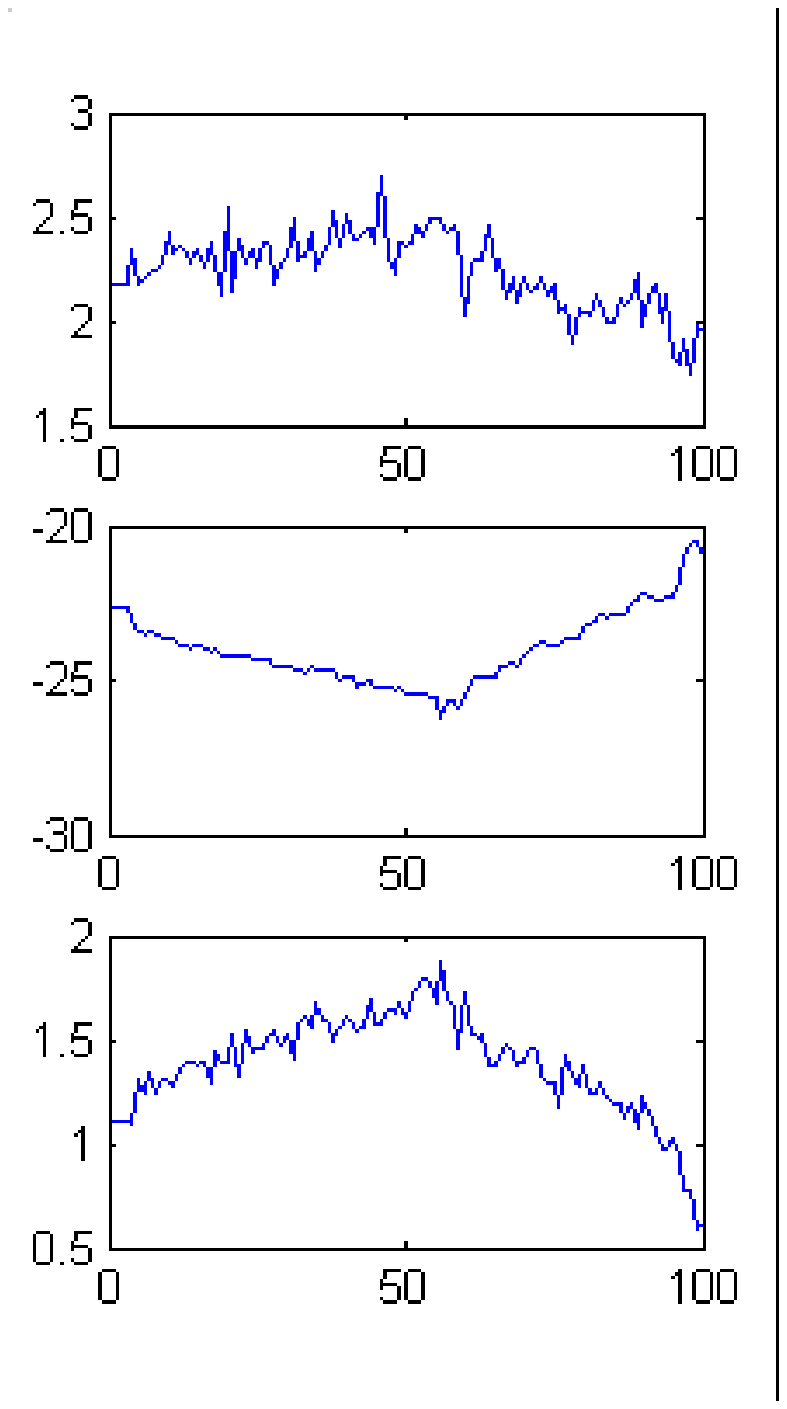}
                 \includegraphics[width=6 true cm,angle=0]{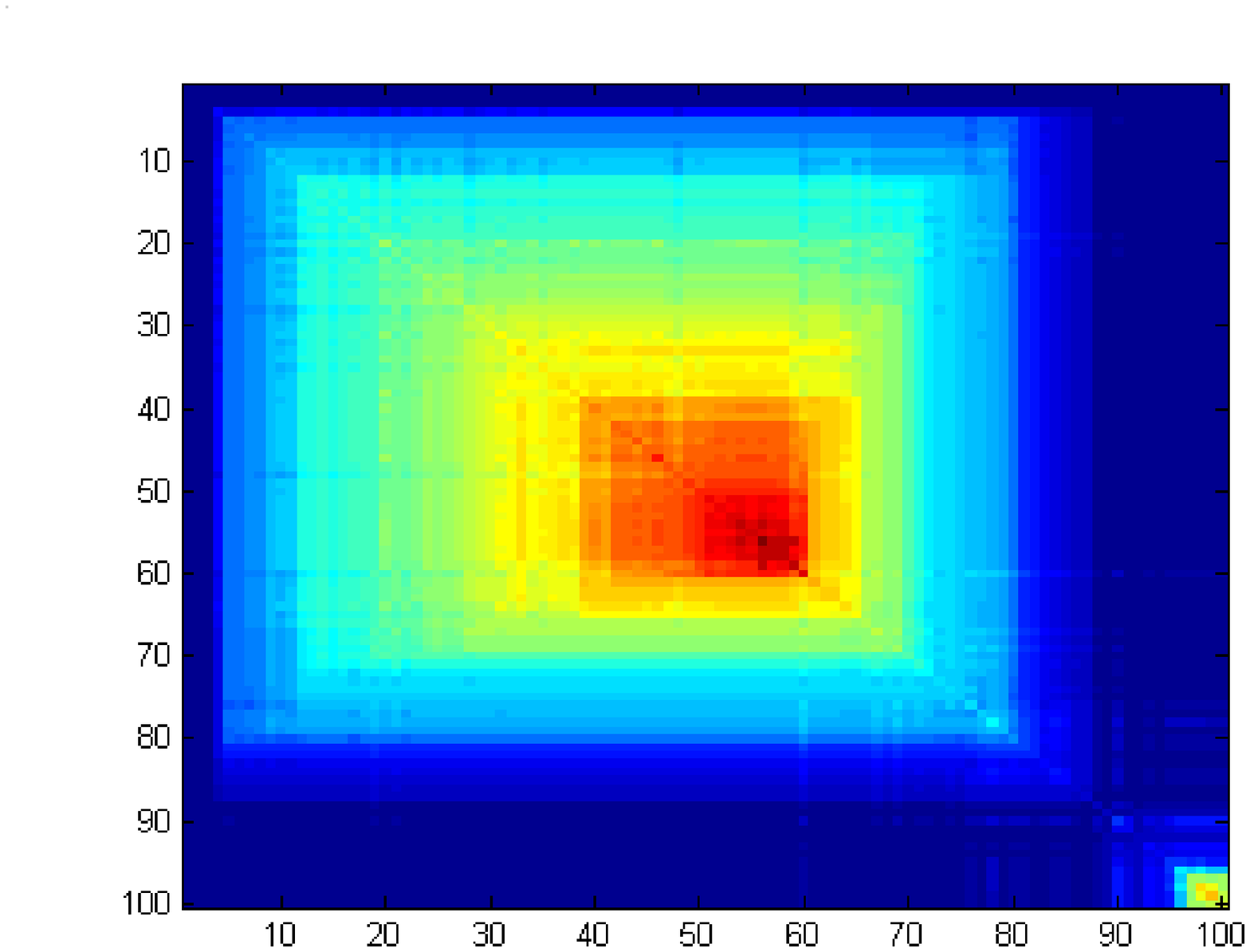}
 \caption{Left: Character values  (top: logs, middle: Kabs, bottom: ldiam) vs taxa arranged according to the best order.  Right: Distance matrix on the 3 characters.}
         \label{fig4}
         \end{center}
 \end{figure}

 \begin{figure}[t]
         \begin{center}
                 \includegraphics[width=8 true cm,angle=0]{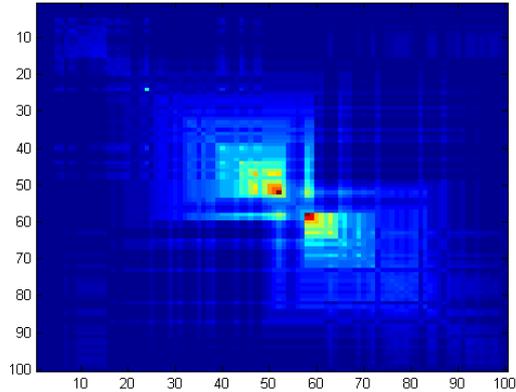}
 \caption{Minimum contradiction matrix for the characters D/B and OIII. }
         \label{fig5}
         \end{center}
 \end{figure}

 \begin{figure}[t]
         \begin{center}
                 \includegraphics[width=10 true cm,angle=0]{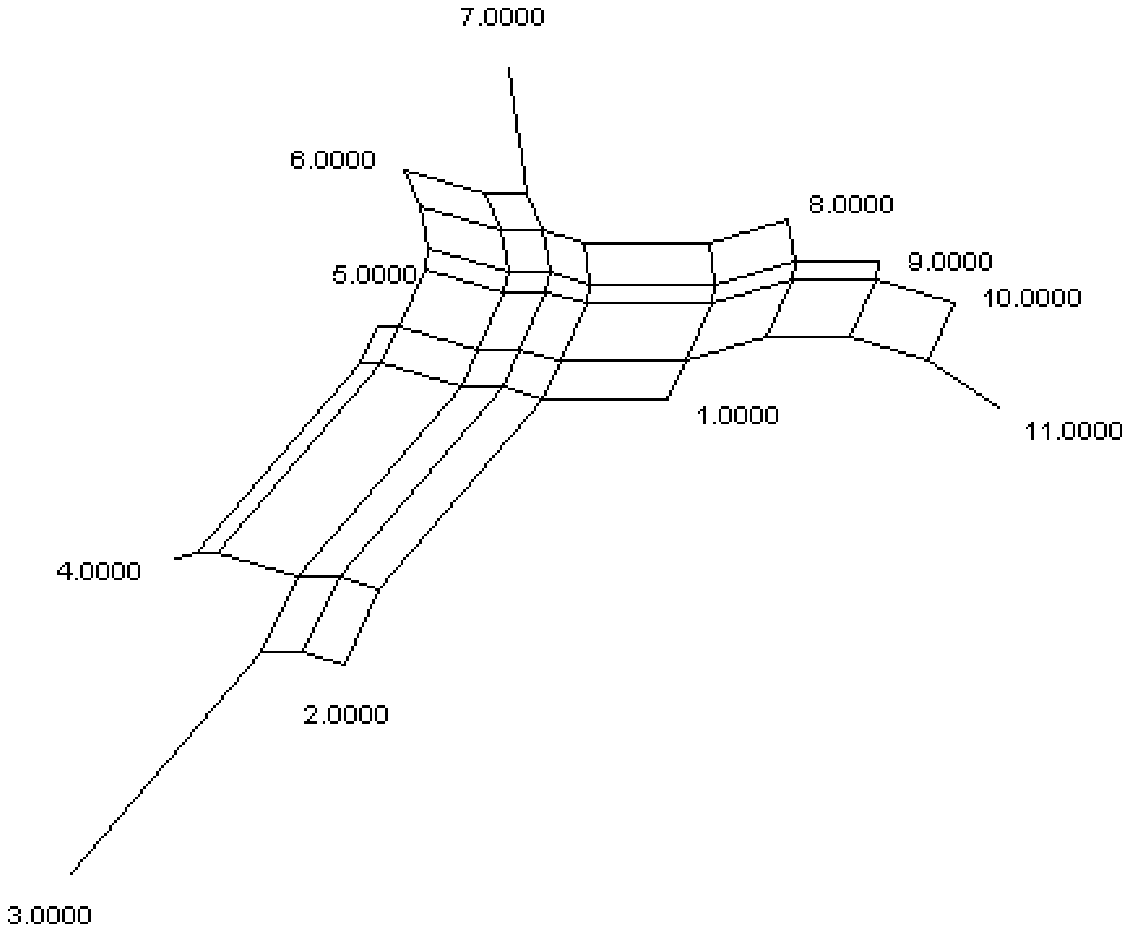}
 \caption{An example of a split network.}
         \label{fig6}
         \end{center}
 \end{figure}

        Cladistics is a difficult concept and its use in the case of continuous variables (morphometric data) is under intense investigation in the bioinformatics community (Gonzàles-Rosé et al 2008). In particular, the formal relationship between distance-based (clustering, continuous variables) and character-based (cladistics, discrete values) is a very interesting topic in itself for which astrophysics brings an entirely new kind of data (Thuillard and Fraix-Burnet 2009). 

        Maximum parsimony and distance-based approaches are the most popular methods to produce phylogenetic trees. While most studies in biology use discrete characters, there is a growing need for applying phylogenetic methods to continuous characters. Examples of continuous data include gene expressions (Planet et al. 2001), gene frequencies (Edwards and Cavalli-Sforza 1964; 1967), phenotypic characters (Oakley and Cunningham, 2000) or some morphologic characters (MacLeod and Forey  2003; González-José et al. 2008). 
The simplest method to deal with continuous characters using maximal parsimony consists of discretizing the variables into a number of states small enough to be processed by the software.

        Distance-based methods are applied to both discrete and continuous input data. Compared to character-based approaches, distance-based approaches are quite fast and furnish in many instances quite reasonable results. As pointed out by Felsenstein (2004), the amount of information that is lost when using a distance-based algorithm compared to a character-based approach is often surprisingly small. The use of continuous characters in distance-based methods may at first glance be less problematic than in character-based methods as algorithms like the Neighbour-Joining work identically on discrete or continuous characters. But also here it is often not easy to determine if the data can be described by a tree. In Thuillard and Fraix-Burnet (2009), we investigate the conditions for a set of continuous characters to describe a split network or a X-tree (split networks are generalized trees). We show that a set of m continuous characters can be described by a split network or a weighted X-tree if the m-dimensional space representation of the taxa state values is on an orthogonal convex hull. In that case, there exists an order of the taxa, called perfect order, for which the distance matrix satisfies the Kalmanson inequalities for each character. 

        In practice, identifying a priori characters that comply to these conditions is difficult. For complex objects in evolution, this requires some good knowledge of the evolution of the characters together with some ideas about the correct phylogeny or at least a rough evolutionary classification. In astrophysics, the study of galaxy evolution has yet not reached this point. However, we can show that the approach presented in this work, based on the Minimum Contradiction Analysis (Thuillard 2007, 2008) is extremely valuable even in cases with very little a priori hints. Here, we illustrate how this method can be used in practice, in particular to discover structuring characters and objects (more can be found in Thuillard and Fraix-Burnet 2009). We have taken from Ogando et al (2008) a sample of 100 galaxies described by some observables and derived quantities. 

        In this first example, three variables have been chosen: logs, that is the logarithm of the central velocity dispersion, Kabs that is the absolute magnitude in the K band and measures the total quantity of near-infrared K light emitted mainly by the stars, and ldiam that is the diameter of the galaxy. They are all expected to be correlated with the total mass of the galaxy. In addition,  Kabs and ldiam both strongly depend on the total number of stars. 

        Figure 4 shows the three variables logs, Kabs, Idiam after ordering of the galaxies with the Minimum Contradiction Analysis that provides the best order (the closest to the perfect order). The two last characters fulfil quite well the conditions for perfect order and the associated distance matrices are well ordered. The first character is in very first approximation well ordered. The distance matrix is well approximated by an expression  characterizing a line tree structure that do not show particular substructures.

        As expected, the variables Kabs and ldiam are well correlated as can be seen from the ordered taxa corresponding to the best order. They are essentially responsible for the two groups seen on the distance matrix of Fig. 4 and thus for the line tree structure. The variable logs is only in first approximation correlated to the other variables. It is much noisier and has no obvious extremum.

        In the second example, we have chosen two variables possibly very little correlated and with evolution function probably not far from the case yielding a convex hull. D/B is the ratio between the size of the disk to that of the bulge. This is a quantitative and objective measure of the morphology of galaxies, spiral (that are disky) systems having a high D/B and elliptical ones having low D/B. The OIII variable is an intensity measure of a region of the spectrum where the OIII emission or absorption line is present. A strong emission line indicates the presence of interstellar gas and thus probably star formation. It is a priori not directly physically linked to the morphology. 

        Despite some large deviations to perfect order for a number of galaxies, the level of contradiction is quite low as can be seen in Figure 5. The distance matrix can be described to a good approximation by a split network. A typical split network is shown in Figure 6.

        Obviously, split networks can represent more complex relationships than trees, but they are very difficult to use for inferring phylogenetic hypotheses, even more than reticulograms that take into account hybridization or horizontal transfer. In Thuillard and Fraix-Burnet (2009), we propose a simple a posteriori discretization of the variables that, in favorable cases, simplifies a split network into a X-tree. This is depicted in Figure 7 where the values of the variables are plotted in the best order of Figure 5.

 \begin{figure}[t]
         \begin{center}
                 \includegraphics[width=10 true cm,angle=0]{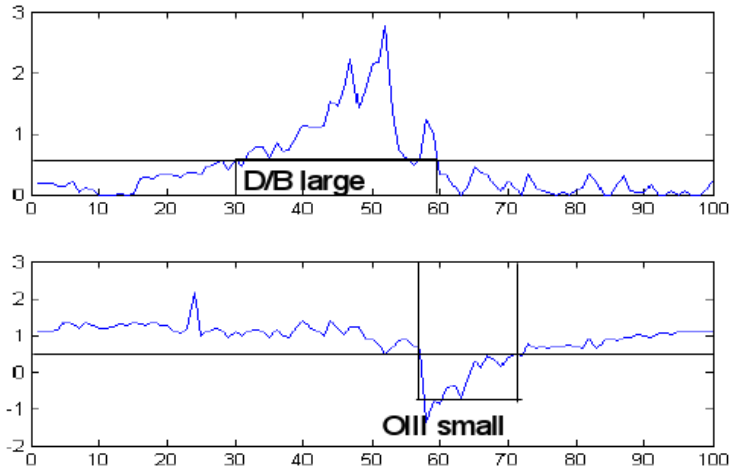}
 \caption{Behavior of the two variables D/B and OIII arranged according to the best order represented in Figure 5. The a posteriori discretization is shown, and yields the tree in Figure 8.}
         \label{fig7}
         \end{center}
 \end{figure}

         Using a threshold value for the two variables OIII and D/B furnishing the discrete characters  OIII large, OIII small, D/B large and D/B small. Compared to the peak in the D/B curve, the large peak for OIII is shifted to the right by roughly 5-10 galaxies. This shift generates some structure in the distance matrix (Figure 5). The galaxies can then be grouped into 3 different groups. The first group has large D/B values and large OIII except for 2 or 3 galaxies that show small OIII, the second group has small D/B and small OIII values while the group 3 contains all galaxies with small values of D/B and large OIII. This grouping can be represented on the tree of Figure 8.

 \begin{figure}[t]
         \begin{center}
                 \includegraphics[width=7 true cm,angle=0]{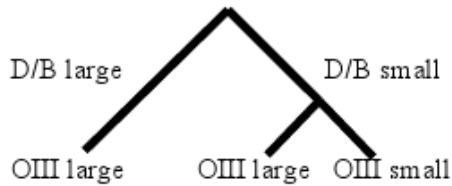}
 \caption{The "phylogenetic" tree obtained from the discretization shown in Figure 7.}
         \label{fig8}
         \end{center}
 \end{figure}

        The first group is made of disky galaxies, which have generally a high star formation rate. There are only 2 or 3 galaxies of this group that have a small OIII. The second and third groups have more bulgy galaxies. In the latter, some galaxies show an OIII line in emission as strong as in group 1. These galaxies seem to contradict the long known observation, that elliptical galaxies have less gas, and thus form less stars than spiral galaxies. This observation is believed to be due to historical reasons in the evolution of galaxies that are not at all clear. Is it due to a sweeping of the gas in the elliptical galaxies by the intergalactic medium? Or is it due to the way elliptical galaxies formed, exhausting nearly all their gas in very efficient star formation episodes? Even if our sample is relatively small, OIII and D/B are obviously not tightly correlated since the first group contains some disky galaxies with low star formation and group 3 has bulgy galaxies with OIII in emission. Anyhow, this example is only for illustration, and does not pretend to generality. At least, it shows that since the formation scenarios for galaxies are complicated, they cannot be studied with two parameters only, like the present morphologies (like D/B) and the OIII signature.

        As a conclusion of this study, the level of contradiction of the minimum contradiction matrix furnishes an objective measure of the  deviations to a tree or network structure. This measure can be used as a criterion to select the most important characters. The minimum contradiction matrix can also be of great help to a posteriori discretize continuous characters so that the resulting discrete characters can be well described by a tree or a split network. Quite interestingly, while discretization of continuous characters is often problematic, the posterior discretization after ordering of the taxa can help removing contradictions from a split network or tree structure. We believe there is here a great hope  not only for astrophysics but for many evolutionary and multivariate studies.

\section{COMMENTS AND CONCLUSION}

        It is now clear that cladistics can be applied and be useful to the study of galaxy diversification. Many difficulties, conceptual and practical, have been solved,. Significant astrophysical results have been obtained and will be extended to larger samples of galaxies and globular clusters. However, many paths remain in the exploration of this new and large field of research.

        There are difficulties that seem to be intrinsic to astrophysics. Most notably, we have millions of objects, but a few tens of descriptors. Of course the situation will improve with time, in particular with integral-field spectroscopy (spectra of detailed regions in the galaxy, e.g. Ensellem et al. 2007). Spectra might not be currently employed at their full capacity of description. Here also, sophisticated statistics can help. But it is not clear whether these data would lead to the discrimination of hundreds of classes. Perhaps this is an erroneous target, perhaps galaxies cannot be classified with such a refinement. Nonetheless, this is already a matter of using multivariate clustering methods and interpreting their results usefully. This is probably more than a conceptual question because clustering with continuous data and large intraspecific variance is a very complicated problem in itself. Sophisticated statistical tools must be used, but the question of characterizing the groups in this context requires a different culture that is not yet widespread in astrophysics. We are convinced that improvements can be made here. 

        Cladistics is supposed to identify clades, that are evolutionary groups, whereas the concept of “species” is not defined at all in astrophysics, and we have even not converged toward groupings based on multivariate analyses. Understanding the formation and evolution of galaxies, depicting a global scenario for galaxy diversification, is a major challenge for the astrophysics of XXIst century.

\section*{REFERENCES}

\begin{enumerate}
 \setlength{\itemsep}{0pt}
   \item Bell, E.F. (2005). Galaxy Assembly. Planets to Cosmology: Essential Science in Hubble's Final Years. Ed. M. Livio. Cambridge: CUP (astro-ph/0408023).
   \item Buchanan, B. and Collard, M. (2008). Phenetics, cladistics, and the search for the Alaskan ancestors of the Paleoindians: a reassessment of relationships among the Clovis, Nenana, and Denali archaeological complexes. Journal of Archaeological Science, 35, 1683-1694.
   \item Cabanac, R.A., de Lapparent, V. and Hickson, P. (2002). Classification and redshift estimation by principal component analysis. Astronomy \& Astrophysics, 389, 1090–1116 (astro-ph/0206062).
   \item Chattopadhyay, T. and Chattopadhyay, A. (2006) Objective classification of spiral galaxies having extended rotation curves beyond the optical radius. The Astronomical Journal, 131, 2452–2468.
   \item Chattopadhyay, T. and Chattopadhyay, A. (2007). Globular Clusters of LOCAL Group – Statistical Analysis. Astronomy \& Astrophysics, 472, 131-140.
   \item Chattopadhyay, T., Misra R., Naskar, M. and Chattopadhyay, A. (2007). Statistical evidences of three classes of Gamma Ray Bursts. Astrophysical Journal, 667, 1017.
   \item Chattopadhyay, T., Mondal, S. and Chattopadhyay, A. (2008). Globular Clusters in the Milky Way and Dwarf Galaxies - A Distribution-Free Statistical Comparison. Astrophysical Journal, 683, 172.
   \item Chattopadhyay, T., Babu, J., Chattopadhyay, A. and Mondal, S. (2009). Horizontal Branch Morphology of Globular Clusters: A Multivariate Statistical Analysis.  Astrophysical Journal, in press.
   \item Edwards, A.W.F. and Cavalli-Sforza, L.L. (1964). Reconstruction of evolutionary trees. Phenetic and Phylogenetic Classification. Ed. V. H. Heywood and J. McNeill. Systematics Association pub. no. 6, London, .pp. 67-76.  
   \item Felsenstein, J. (2004). Inferring phylogenies, Sinauer Associates.
   \item Fraix-Burnet, D., Choler, P., Douzery, E. and Verhamme, A. (2006a). Astrocladistics: a phylogenetic analysis of galaxy evolution. I. Character evolutions and galaxy histories. Journal of Classification, 23, 31-56 (astro-ph/0602581).
   \item Fraix-Burnet, D., Choler, P. and Douzery, E. (2006b). Towards a Phylogenetic Analysis of Galaxy Evolution: a Case Study with the Dwarf Galaxies of the Local Group. Astronomy \& Astrophysics, 455, 845-851 (astro-ph/0605221).
   \item Fraix-Burnet, D., Davoust, E. and Charbonnel, C. (2009). The environment of formation as a second parameter for globular cluster classification. Monthly Notices of the Royal Astronomical Society, 398, 1706-1714.
   \item Fraix-Burnet, D., Douzery, E., Choler, P. and Verhamme, A. (2006c). Astrocladistics: a phylogenetic analysis of galaxy evolution. II. Formation and diversification of galaxies. Journal of Classification, 23, 57-78 (astro-ph/0602580).
   \item Fraix-Burnet D. (2009). Galaxies and Cladistics. Evolutionary Biology: Concept, Modeling, and Application. Ed. Pontarotti P. Biomedical and Life Sciences, Springer Berlin Heidelberg, pp. 363–378 \\ (http://fr.arxiv.org/abs/0909.4164).
   \item Gonzàlez-José, R., Escapa, I., Neves, W.A., Cúneo, R. and Pucciarelli, H.M. (2008). Cladistic analysis of continuous modularized traits provides phylogenetic signals in Homo evolution. Nature, 453, 775-779.
   \item Hernandez, X.and Cervantes-Sodi, B. (2005). A dimensional study of disk galaxies.  11th Latin-American Regional IAU Meeting (astro-ph/060225)
   \item MacLeod, N. and Forey, P.L. (2003). Morphology, Shape and Phylogeny. Eds. Taylor and Francis Inc., New York.
   \item Neistein, E., van den Bosch, F.C. and Dekel, A. (2006). Natural downsizing in hierarchical galaxy formation. Monthly Notices of the Royal Astronomical Society,  372, 933-948 (astro-ph/0605045).
   \item Oakley, T.H. and Cunningham, C.W. (2000). Independent contrasts succeed where ancestor reconstruction fails in a known bacteriophage phylogeny. Evolution, 54 (2), 397-405.
   \item Ogando, R.L.C., Maia, M.A.G., Pellegrini, P.S. and da Costa, L.N. (2008). Line strengths of early-type galaxies. The Astronomical Journal, 135, 2424-2445 (http://fr.arxiv.org/abs/0803.3477).
   \item Planet, P.J, DeSalle, R., Siddal, M., Bael, T., Sarkar, I.N. and Stanley, S.E. (2001). Systematic analysis of DNA microarray data: ordering and interpreting patterns of gene expression . Genome Research, 11, 1149-1155.
   \item Recio-Blanco, A., Aparicio, A., Piotto, G., De Angeli, F. and Djorgovski, S.G. (2006). Multivariate analysis of globular cluster horizontal branch morphology: searching for the second parameter. Astronomy \& Astrophysics, 452, 875-884 (http://arxiv.org/abs/astro-ph/0511704).
   \item Thuillard, M. (2007). Minimizing contradictions on circular order of phylogenic trees. Evolutionary Bioinformatics, 3, 267-277.
   \item Thuillard, M. (2008).  Minimum contradiction matrices in whole genome phylogenies. Evolutionary Bioinformatics, 4, 237-247.
   \item Thuillard M. and Fraix-Burnet D. (2009). Phylogenetic Applications of the Minimum Contradiction Approach on Continuous Characters. Evolutionary Bioinformatics, 5, 33-46 (http://fr.arxiv.org/abs/0905.2481).
   \item Wiley, E.O., Siegel-Causey, D., Brooks, D.R. and Funk, V.A.  1991). The Compleat Cladist: A Primer of Phylogenetic Procedures. The university of Kansas, Museum of Natural History special publication N°19.
\end{enumerate}

\end{document}